\documentstyle[aps,prd,twocolumn,epsf,epsfig]{revtex}

\newcommand \bea{\begin{eqnarray}}
\newcommand \eea{\end{eqnarray}}
\newcommand \ga{\raisebox{-.5ex}{$\stackrel{>}{\sim}$}}
\newcommand \la{\raisebox{-.5ex}{$\stackrel{<}{\sim}$}}

\begin{document}
\twocolumn[\hsize\textwidth\columnwidth\hsize
\csname@twocolumnfalse%
\endcsname
\draft
\title{Collective modes of trapped gases at the BEC-BCS crossover}
\author{H. Heiselberg}
\address{Danish Defense Research Establishment, Ryvangsalle' 1, 2100 Copenhagen \O, Denmark}
\maketitle
\begin{abstract}
The collective mode frequencies in isotropic and deformed traps are
calculated for general polytropic equation of states, $P\propto
n^{\gamma+1}$, and expressed in terms of $\gamma$ and the trap geometry.
For molecular and standard Bose-Einstein condensates and Fermi gases
near Feshbach resonances, the
effective power $\gamma\simeq0.5-1.3$ is calculated from Jastrow type
wave-function ans\"atze, and from the crossover model of Leggett. 
The resulting mode frequencies are
calculated for these phases around the BCS-BEC crossover.
\\ 
\end{abstract}
\vskip1pc]

Recent experiments probe systems of fermions
\cite{Thomas,ENS,Regal,Ketterle,Innsbruck} and bosons \cite{Claussen}
near Feshbach resonances by expansion and RF spectroscopy.
Interesting new strongly interacting or dense phases of bosons and
fermions are created, e.g., that associated with the crossover from a
(possibly superfluid) Fermi gas to a molecular BEC. The corresponding
equations of states (EOS) differ from standard dilute systems which
directly shows up in their collective modes.

The purpose of this work is to calculate the 
collective modes in terms of a general class
of polytropic EOS, to calculate the EOS for strongly interacting
BEC and Fermi gas at the crossover to a molecular BEC, and finally synthesize
the two to calculate the collective modes for these strongly interacting 
phases. 

Polytropic EOS relate the pressure and density as
\bea
   P \propto n^{\gamma+1} \,.
\eea
As we shall show below
the collective modes in harmonic oscillator traps
depend on the power $\gamma$ but not on other details of the polytropic EOS. 
Polytropic EOS apply to many systems.
In a dilute interaction dominated BEC $\gamma=1$,
whereas an ideal Bose gas in the
normal state has $\gamma=2/3$ under adiabatic conditions.
A dilute gas of Fermi atoms also has  $\gamma=2/3$ in both
the hydrodynamic and superfluid limits. 
Both a Fermi gas \cite{HH} and a BEC \cite{Cowell}
has $\gamma=2/3$ in the strongly interacting (unitarity) limit.
We shall see below that near a Feshbach resonance, where a
Fermi gas crossover to a molecular BEC,
the power effectively varies between $\gamma\sim 0.5-1.3$.

The collective modes are calculated from the equations of motion
which in hydrodynamics and for a superfluid are given by the
equation of continuity and the Euler equation
\bea
   mn\frac{\partial{\bf v}}{\partial t} =
   -\nabla P -n\nabla V_{ext} \,.
\eea
Here $n$ and ${\bf v}$ are the local density and velocity, and 
$V_{ext}=(1/2)m\sum_i \omega_i^2r_i^2$ is the harmonic oscillator trap 
potential.
From the Euler equation we obtain the equilibrium density:
$n_{eq}=n_0(1-\sum_i r_i^2/R_i^2)^{1/\gamma}$,
where $R_i^2=2(\gamma+1)P_0/\gamma n_0m\omega_i^2$, $i=1,2,3$, are the 3D 
Thomas-Fermi radii of the trapped cloud of atoms ($P_0$ and $n_0$ are the
pressure and density in the center of the trap). 

Linearizing around equilibrium, $n=n_{eq}+e^{i\omega t}\delta n$,
the equations of motion lead to
\bea \label{EOM}
   -m\omega^2 \delta n = \nabla\cdot 
     \left[n\nabla\left(\frac{1}{n}\frac{dP}{dn}\delta n\right)\right] \,.
\eea
It is not necessary to restrict ourselves to zero temperature where the 
Gibbs-Duhem relation $dP=nd\mu$ simplifies Eq. (\ref{EOM}).

In an isotropic trap the collective modes with angular momentum $l$ and
$n$ radial nodes are straight forward to calculate from
Eq. (\ref{EOM}) by generalizing the method of Ref. \cite{PS} to any
polytropic EOS. We find that the departure from the equilibrium
density is $\delta n\propto
r^l(1-r^2/R^2)^{(1/\gamma-1)}F(-n,n+l+\gamma^{-1},l+3/2,r^2/R^2)$,
where $F$ is the hypergeometrical function. The corresponding
eigenvalues are
\bea \label{spherical}
  \frac{\omega^2}{\omega_0^2} = l+2n[\gamma(n+l+1/2)+1] \,,
\eea  
which reduces to the known results for $\gamma=1$ \cite{Stringari} and
$\gamma=2/3$ \cite{Baranov}.
In comparison the collective modes in the collisionless limit are those of a
free particle: $\omega/\omega_0=2n+l$, when its mean free path exceeds the
size of the cloud.

The hydrodynamic collective modes can also be calculated for deformed traps
for a general polytropic EOS. 
Linearizing the equations of motion lead to the following equation for the
collective modes
\bea 
  -\omega^2{\bf v} = \nabla({\bf v}\cdot \nabla V_{ext}) + 
   \gamma(\nabla V_{ext}){\bf (\nabla\cdot v)} \,.
\eea
The breathing modes have flow velocity on the form 
${\bf v}=(a_1r_1,a_2r_2,a_3r_3)e^{i\omega t}$, which leads
to three coupled homogeneous equation for $a_{i=1,2,3}$
\bea
  \gamma \sum_j \omega_j^2a_j = (\omega^2-(\gamma+2)\omega_i^2)a_i \,.
\eea
In an axial symmetric trap: 
$\omega_1=\omega_2\equiv\omega_0$ and $\omega_3=\lambda\omega_0$,
the resulting breathing modes are \cite{Cozzini}
\bea \label{deformed}
   \frac{\omega^2}{\omega_0^2} &=&\gamma+1+\frac{\gamma+2}{2}\lambda^2 
    \nonumber\\    &\pm& 
  \sqrt{(\gamma+2)^2\lambda^4/4+(\gamma^2-3\gamma-2)\lambda^2+(\gamma+1)^2}
 \,,
\eea
and $\omega=\sqrt{2}\omega_0$.
The $\pm$ eigenvalues are the radial and axial modes respectively and result
from the coupled monopole and quadrupole $m=0$ modes, where
$m$ is the angular momentum projection on the 3rd axis.
Therefore, the breathing modes for an isotropic trap $\lambda=1$ become
the quadrupole with $\omega=\sqrt{2}\omega_0$ and the monopole with
$\omega=\sqrt{3\gamma+2}\omega_0$ as follows from both
Eq. (\ref{deformed}) and Eq.  (\ref{spherical}) for $n=0,l=2$ and
$n=1,l=0$ respectively.  For $\gamma=1$ and $\gamma=2/3$ the monopole
frequencies are the standard $\omega=\sqrt{5}\omega_0$ and
$\omega=2\omega_0$ respectively.

For a very elongated or cigar-shaped trap (prolate in nuclear terminology), 
$\lambda\ll 1$, used in recent experiments \cite{Innsbruck}, 
Eq. (\ref{deformed}) results in a low frequency axial mode with
\bea \label{ax}
  \omega_{ax} = \sqrt{3-(\gamma+1)^{-1}}\, \omega_3 \, .
\eea
For $\gamma=1$ and $\gamma=2/3$ the
axial mode frequencies are  $\omega=\sqrt{5/2}\omega_3$ and
$\omega=\sqrt{12/5}\omega_3$ respectively as found in \cite{Pit,Amo}.
The radial modes have
\bea \label{rad}
  \omega_{rad} = \sqrt{2(\gamma+1)} \,\omega_0 \,.
\eea
For $\gamma=1$ and $\gamma=2/3$ the
radial mode frequencies are the standard $\omega=2\omega_0$ and
$\omega=\sqrt{10/3}\omega_0$ respectively.

In the oblate limit, $\lambda\gg 1$, the breathing modes
are $\omega_{ax}=\sqrt{\gamma+2}\omega_3$ and
$\omega_{rad}=\sqrt{(6\gamma+4)/(\gamma+2)}\omega_0$. They connect to the
prolate limit through avoided level crossing as seen in Fig. 1.

\vspace{-0.5cm}
\begin{figure}
\begin{center}
\psfig{file=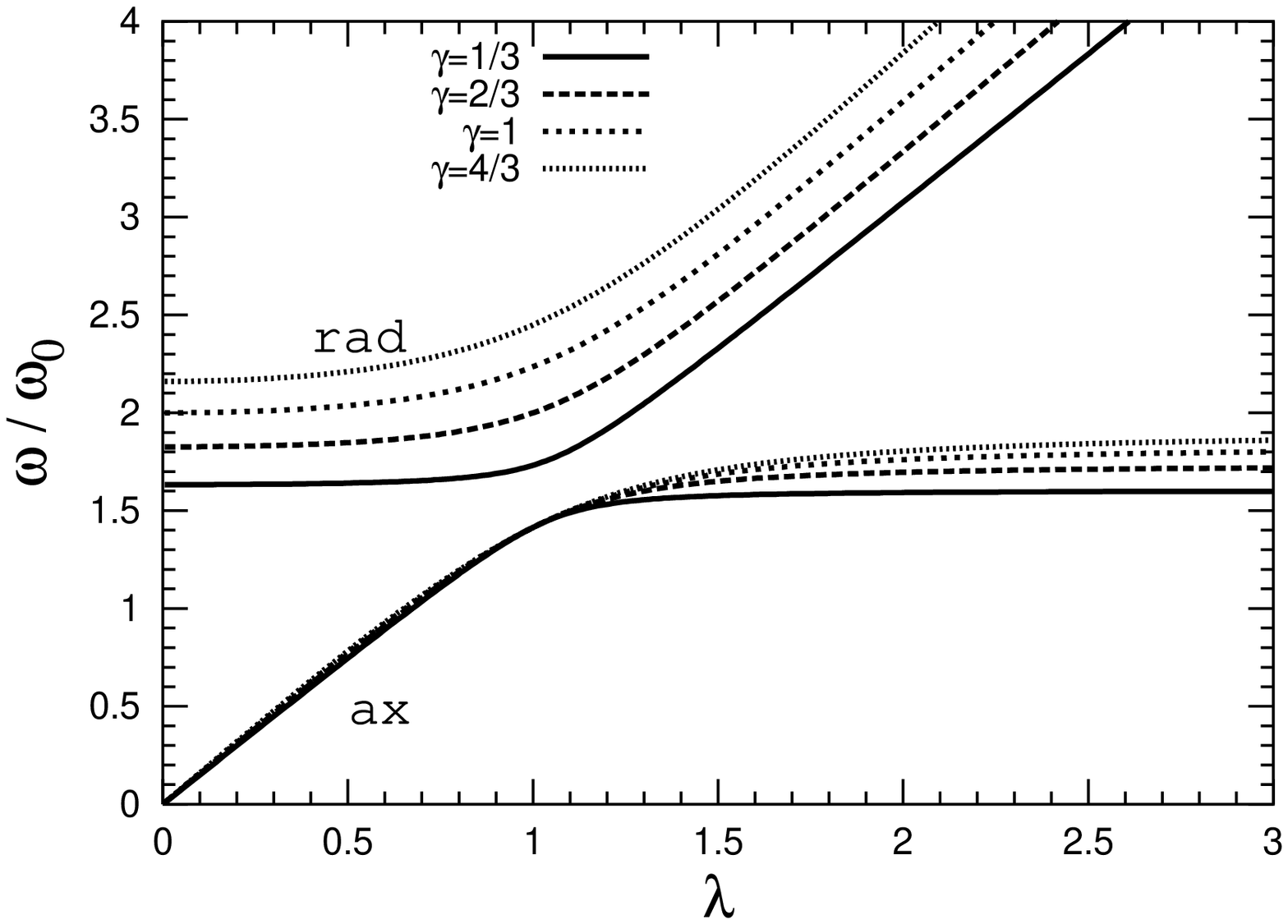,height=8.0cm,angle=-90}
\vspace{.2cm}
\begin{caption}
{The  $\pm$ solutions of Eq. (\ref{deformed}) vs. trap deformation
$\lambda=\omega_3/\omega_0$ for $\gamma=1/3,2/3,1,4/3$.
For a cigar shaped trap, $\lambda\ll1$, these correspond to the
radial and axial mode frequencies respectively.
}
\end{caption}
\end{center}
\label{f1}
\end{figure}
\vspace{-.2cm}

We now turn to the EOS for strongly interacting Bose and Fermi gases and
calculate an effective polytropic index that can be applied for the above
modes. The EOS for a BEC was calculated in Ref. \cite{Cowell} from a 
Jastrow type wave function 
$\Psi_{J}({\bf r}_1,...,{\bf r}_N)=\prod_{i<j}f({\bf r}_i-{\bf r}_{j})$,
which incorporates essential two-body correlations and is a good
approximation for cold dilute and dense bose systems \cite{Vijay}.  
It was shown that with proper boundary conditions the calculated energy
reproduced the dilute limit result,
$E/N = 2\pi \hbar^2 an/m$, 
where $a$ is the s-wave scattering length between bosons.
However, in the unitarity limit,
$n^{1/3}a\gg 1$ or $x=1/ak_F\simeq 0$, the energy per particle scales
like a Fermi gas polytrope:
$E/N = 13.33 \hbar^2 n^{2/3}/m=2.79E_F$. 
Here, we have also for bosons defined $E_F=\hbar^2k_F^2/2m$ in terms of 
the density $n=k_F^3/3\pi^2$ for later comparison between molecular BEC
and Fermi gases with two spin states.
In this model (see \cite{Cowell,Levico} and Fig. 2 for details)
we can calculate the zero temperature pressure $P=n^2d(E/N)/dn$
and the effective polytropic index, which we define as
the logarithmic derivative
\bea \label{gamma}
    \bar{\gamma} \equiv \frac{n}{P}\frac{dP}{dn} \, -1\, .
\eea
As shown in Fig. 3  $\bar{\gamma}$ approaches
1 asymptotically from above in the dilute limit but is 2/3 in the 
unitarity limit with a maximum of $\sim$1.25 around $1/k_Fa_m\sim 1.4$.

\vspace{-0.5cm}
\begin{figure}
\begin{center}
\psfig{file=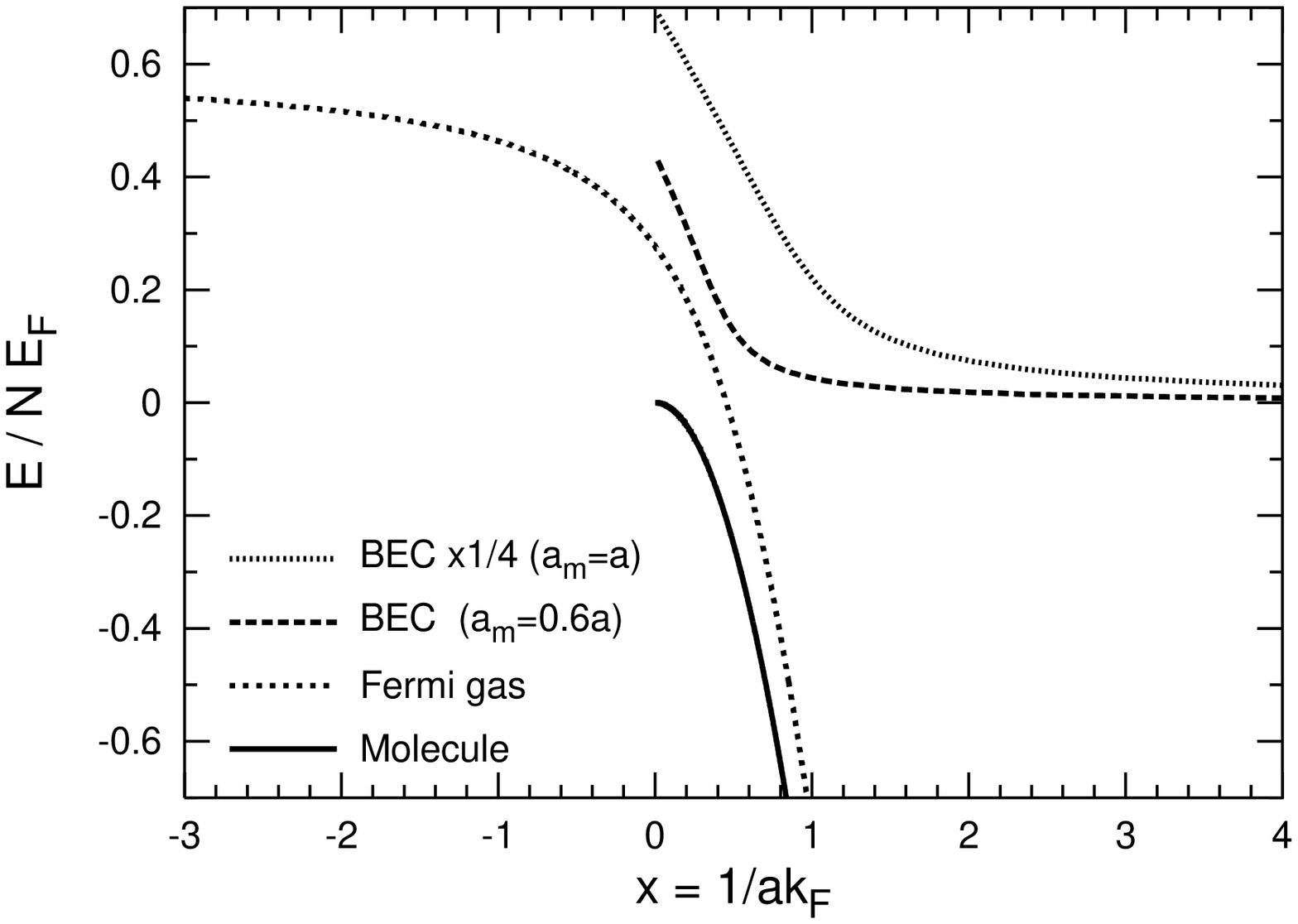,height=8.0cm,angle=-90}
\vspace{.2cm}
\begin{caption}
{The energy per particle in units of $E_F$ for a BEC with $a_m=a$ and
$a_m=0.6a$, and a Fermi gas as it crossover towards a molecular
BEC (see
text). At small positive scattering length the energy per fermion
approaches half the binding energy of a molecule, $E/N\to
-\hbar^2/2ma^2=-x^2E_F$.
}
\end{caption}
\end{center}
\label{f2}
\end{figure}
\vspace{-.2cm}

 The approximate EOS can also be applied to a molecular BEC when a
number of factors are taken into account: the density of molecules is
half that of Fermi atoms, their mass is two times larger, the above
calculated $E/N$ is for two atoms, and finally the scattering length
between molecules ($a_m$) may differ from that between the two Fermi
atoms forming the molecule. Petrov et al. find \cite{Petrov}
$a_m=0.6a$ in accordance with \cite{Regal}.
As a result the energy in molecular BEC is per atom much smaller than that of
a BEC of Bose atoms as seen in Fig. 2. 
This is important when we attempt to match 
the EOS of a molecular BEC onto a
gas of Fermi atoms in the unitarity limit.
The difference in magnitude cancels in Eq. (\ref{gamma}) and
is therefore not important for calculating $\bar{\gamma}$. 
A difference between $a$ and $a_m$ and between Bose and Fermi densities
does, however, affects $ak_F$ as seen in Fig. 3.

The EOS for a Fermi gas at zero temperature near a Feshbach resonance and its
crossover to a molecular BEC has recently been studied by various
resummation techniques \cite{Baker,HH}, a Jastrow-Slater 
type ansatz \cite{HH,Levico}
and by fixed-node Greens function Monte Carlo (FN-GFMC) \cite{Carlson}.
The EOS calculated from 
the Jastrow-Slater ansatz has the merit that the EOS is exact to leading 
orders both
in the dilute and molecular limits. Furthermore, it has been tested
experimentally \cite{Levico} and in FN-GFMC to be a good approximation in
the unitarity limit as well.  
It extends the Jastrow wave function for bosons described above by
including an anti-symmetric Slater wave function $(\Phi_S)$, 
which is the ground state of free fermions
$\Psi_{JS}({\bf r}_1,...,{\bf r}_N)= 
\Phi_S\prod_{i<j'}f({\bf r}_i-{\bf r}_{j'})$.
Because $\Phi_S$ insures that same spins are spatially anti-symmetric, 
the Jastrow wave function only applies to particles with different spins
(indicated by the primes). 
The resulting energy (see \cite{Carlson,Levico} for details)
is shown in Fig. 2.
The energy per particle at zero temperature
can generally be written in terms of the ratio between
the interaction and kinetic energies $\beta=E_{int}/E_{kin}$ as 
\cite{HH,Thomas}
\bea
  E/N = E_{kin}+E_{int} = \frac{3}{5}E_F [1+\beta] \,.
\eea
As seen in Fig. 2 it approaches the ideal Fermi gas result
$(3/5)E_F$ in the dilute limit,
$(3/5)(1+\beta)E_F$ in the unitarity limit with $\beta\simeq -0.54$
for two spin states \cite{Levico}, 
and $E/N=-\hbar^2/2ma^2$ in the molecular BEC limit ($a\to 0_+$).
The corresponding pressure is
\bea \label{PF}
   P = n^2\frac{dE/N}{dn}
     =\frac{2}{5} E_F n[ 1+\beta-x\beta'/2] \,.
\eea
Here, we view $\beta(x)$ as a smooth function
of $x=1/ak_F$ with derivatives $\beta'=d\beta(x)/dx$, etc. 
The effective polytropic index is from Eqs. (\ref{gamma}) and (\ref{PF})
\bea \label{gammaF}
    \bar{\gamma} 
    =  \frac{\frac{2}{3}(1+\beta) -x\beta'/2
     +x^2\beta''/6}{1+\beta-x\beta'/2} \,.
\eea
When $\beta(x)$ is a smooth function at $x=0$
we find from Eq. (\ref{gammaF}) that  $\bar{\gamma}=2/3$
in the unitarity limit.

The effective polytropic index is shown in Fig. 3 for a Fermi
gas as it crossover to a molecular BEC. That it turns over and drops
back to $\gamma\to 2/3$ for $x\ga 0.5$ is an artifact of the EOS
resulting from the Slater ansatz in the wave function. The true ground
state wave function is expected to have a lower energy in the
molecular BEC limit as $a\to 0$ as is also found in FN-GFMC
\cite{Carlson}. The Jastrow part of the wave function is responsible
for the correct leading part of the energy: $E/N=-\hbar^2/2ma^2$
which, however, does not contribute to the pressure because it is
density independent. The Slater wave function is responsible for the
leading density dependent order but detailed comparison to FN-GFMC
calculations show that it is only correct up to $x\la 0.5$. In FN-GFMC
a better ground state wave function is found numerically which has
lower energy. Both Fermi gases and BEC's have
$\gamma=2/3$ at $x=0$ due to the universal scaling law $E/N\propto
n^{2/3}$ in the unitarity limit \cite{HH,Cowell}.

Leggett \cite{Leggett} extended the BCS gap equation to the BCS-BEC
crossover and calculated the gap and chemical potential. From the Gibbs-Duhem
relation and Eq. (\ref{gamma}) we
can then calculate $\bar{\gamma}$ as shown in Fig. 3.
In the dilute BCS limit it differs from the Fermi gas, which has
chemical potential $\mu=E_F+2\pi \hbar^2an/m$, by lacking this second
term proportional to $a$. In the other limit the chemical potentials
differ for orders higher than linear in the scattering length. In both
the dilute BEC and the Jastrow BEC approximation higher orders add
positively - which is responsible for $\bar{\gamma}>1$ for $k_Fa_m\ga
1$. The Legget model leads to negative higher order
contributions beyond the linear one: $\mu=-\hbar^2/2ma^2+\pi\hbar^2 a
n/m$, and therefore $\bar{\gamma}$ decreases monotonously from 1
towards 2/3 as $x\to 0_+$.

In all the above models the two-body
correlation function undergoes a smooth transition from a constant in
the dilute Fermi gas limit to that of a molecule in the dilute
molecular Bose gas limit, and the BCS-BEC crossover is continuous. In
the unitarity region the correlation length is of order the
interparticle spacing $\sim k_F^{-1}$; only in the dilute BEC limit
$x\gg 1$ is the correlation length sufficiently small that the
molecules may be approximated as point particles.

\vspace{-0.5cm}
\begin{figure}
\begin{center}
\psfig{file=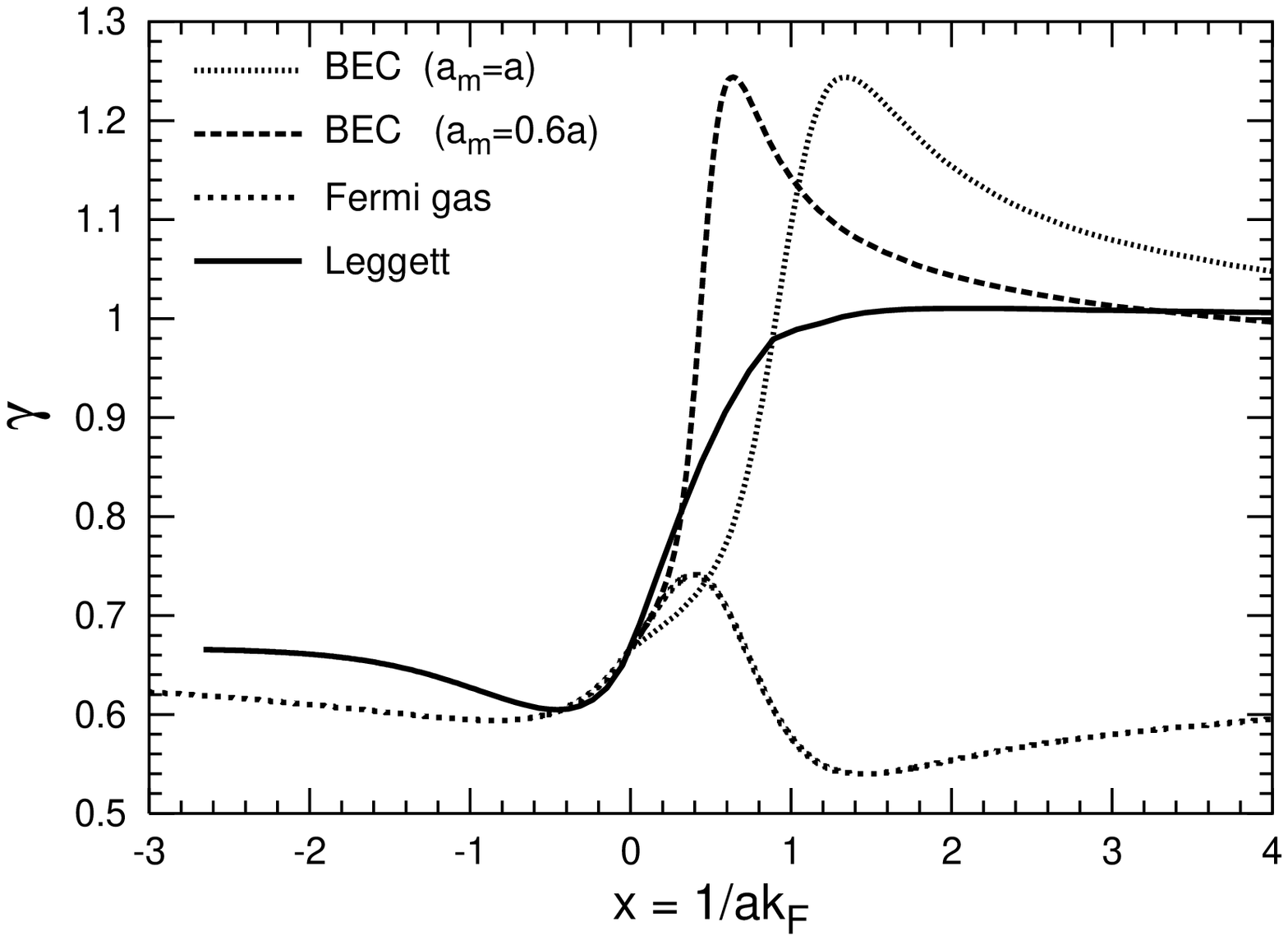,height=8.0cm,angle=-90}
\vspace{.0cm}
\begin{caption}
{The polytropic power $\bar{\gamma}$ 
of a BEC with $a_m=a$ and $a_m=0.6a$,
a Fermi gas (Jastrow-Slater ansatz), 
and the Leggett model describing the smooth crossover to 
a molecular BEC. See text and Fig. 2.}
\end{caption}
\end{center}
\label{f3}
\end{figure}
\vspace{-.2cm}

The strongly interacting EOS's near the 
unitarity limits can be approximated by a polytrope by replacing $\gamma$ with
$\bar{\gamma}$. This allows us to calculate the
collective modes directly from Eqs. (\ref{spherical}) and (\ref{deformed})
as shown in Fig. 4.
Experimentally one tunes the
scattering length near a Feshbach resonance for a fixed number of trapped
particles $N$ whereby the size and density of the cloud and therefore
also $k_F$ varies in a complicated way depending on the EOS. 
For fermions in the dilute and unitarity limit
$\bar{\gamma}=2/3$ and the size of cloud is 
$R=(24N)^{1/6}a_{osc}(1+\beta)^{1/4}$ with $\beta=0$ and $\beta\simeq -0.56$
respectively. Since $\bar{\gamma}\simeq 2/3$ for
a Fermi gas up to and around the unitarity limit this relation for the size
is a good approximation in this region, and  analogously:
$k_F\simeq(24N)^{1/6}a_{osc}^{-1}(1+\beta)^{1/4}$.
In a dilute BEC: $R=(15Na_m)^{1/5}a_{osc}^{4/5}$ and 
$k_F=(3\pi/8)^{1/3}(15N)^{2/15}a_m^{-1/5}a_{osc}^{-4/5}$ in the center of the
trap.

The resulting collective modes - in particular the radial breathing
frequency - are as shown in Fig. 4 very sensitive to $x=1/ak_F$
through $\gamma$ near the unitarity limit, where the Fermi gas crossover to a
molecular BEC. The modes can therefore be exploited to extract the EOS
experimentally. The leading corrections in the dilute Fermi gas and BEC
were given in \cite{String}. It is observed from Fig. 4 that the
frequencies are very sensitive to $a_m/a$ and may therefore be
exploited to relate the atomic and molecular scattering lengths.

\vspace{-0.5cm}
\begin{figure}
\begin{center}
\psfig{file=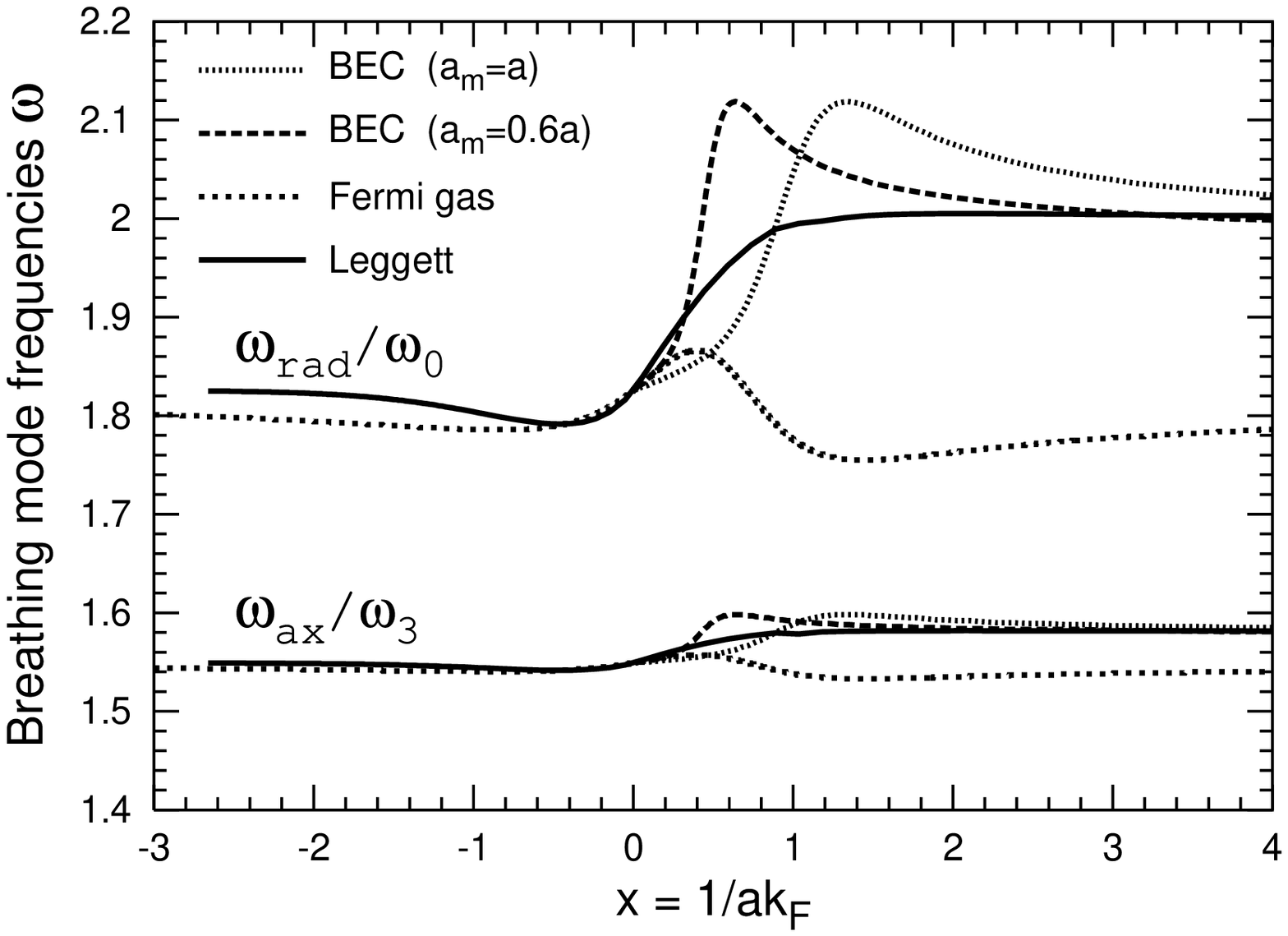,height=8.0cm,angle=-90}
\vspace{.0cm}
\begin{caption}
{The radial and axial frequencies for cigar shaped traps, $\lambda\ll 1$,
from Eqs. (\ref{ax}) and (\ref{rad}) with $\gamma$ from from Fig. 3.}
\end{caption}
\end{center}
\label{f4}
\end{figure}
\vspace{-.2cm}

In very dilute Fermi gases, where the pairing gap becomes smaller than
the oscillator frequency and the coherence length exceeds the system
size, particle excitations of order $\sim 2\Delta$ appear
\cite{Baranov,BruunMottelson} besides the collective modes described
above.  Using the pairing gap in an isotropic dilute trap
\cite{BruunMottelson} the condition $\Delta<\hbar\omega_0$ becomes
$x=1/ak_F\la -(2/\pi)(C+\ln(3N)/3)$, where $C=0.577..$ is Eulers
constant.  Such pair excitation modes are therefore only observable
for weak attractions.

The collective mode frequencies do not distinguish between a
superfluid and a hydrodynamic Fermi gas. The damping of the modes
should be different but has
not been estimated in the unitarity limit for bosons or fermions.
Assuming an unitarity limited scattering cross section we expect the
collision rate to decrease as $\sim exp(-\Delta(T)/T)$ at
temperatures well below the gap $\Delta(T=0)\simeq 0.54E_F\exp(\pi x/2)$ 
\cite{Carlson,Levico} in a bulk system. 
The damping can
potentially discriminate between hydrodynamic and superfluid Fermi
gases, and at the same time the mode frequency discriminates
collisionless.

In summary, the dependence of collective mode frequencies and damping
on density, interaction strength and temperature as described above
can - especially near the BCS-BEC crossover - reveal the underlying
EOS in detail including possible phase transitions and associated critical
temperatures and densities.

{\it Note added in proof}: Two experiments have recently published the
axial and radial modes around the unitarity limit. The above results
for the Leggett model is in nice agreement with the data of Kinast et
al. \cite{Kinast} and (for the axial mode) with Bartenstein et
al. \cite{Bartenstein}.  See also \cite{Hu} for details. The Leggett model
also explains ``surprise one'' in \cite{Bartenstein}


\end{document}